Title: Lax Pair for Strings in Lunin-Maldacena Background
Authors: Sergey Frolov
Comments: 28 pages, v2: a discussion of a 4+2 parameter solution added in section 5;
new Appendix B with a 6+2 parameter deformation of AdS_5 x S^5;
typos corrected, references added.

Recently Lunin and Maldacena used an SL(3,R) transformation of the
AdS_5 x S^5 background to generate a supergravity solution
dual to a so-called beta-deformation of N = 4 super Yang-Mills
theory. We use a T-duality-shift-T-duality (TsT) transformation to obtain
the beta-deformed background for real beta, and show that solutions
of string theory equations of motion in this background are in one-to-one
correspondence with those in  AdS_5 x S^5 with
twisted boundary conditions imposed on the U(1) isometry fields.
We then apply the TsT transformation to derive a local and periodic Lax pair for
the bosonic part of string theory in the beta-deformed background. We also
perform a chain of three consecutive TsT transformations to generate
a three-parameter deformation of AdS_5 x S^5. The three-parameter background
is dual to a nonsupersymmetric marginal deformation of N=4 SYM.
Finally, we combine the TsT transformations with SL(2,R) ones to obtain
a 6+2 parameter deformation of AdS_5 x S^5.

 \documentclass[12pt]{article}
\usepackage{amsmath}
\usepackage{graphicx}
\usepackage{amsfonts}
\usepackage{amssymb}

\begin{document}
\input epsf

\def \tg {\tilde \gamma}
\def\tb{\tilde{\beta}}
\def\l{\lambda}
\def\tl{\tilde{\l}}
\def\z{\zeta}
\def\s{\sigma}
\def\ts{\tilde{\sigma}}
\def\tm{\tilde{\mu}}
\def\g{\gamma}
\def\tg{\tilde{\gamma}}

\def\ttbp{\tilde{\tilde{{\bold p}}}}
\def\bp{{\bold p}}
\def\={\, =\, }
\def\+{\, +\, }
\def\-{\, -\, }
\def\half{{\textstyle{1\over2}}}
\let\a=\alpha \let\b=\beta \let\g=\gamma \let\d=\delta \let\e=\epsilon
\let\z=\zeta \let\h=\eta \let\th=\theta \let\i=\iota \let\k=\kappa
\let\l=\lambda \let\m=\mu \let\n=\nu \let\x=\xi \let\p=\pi \let\r=\rho
\let\s=\sigma \let\t=\tau \let\u=\upsilon \let\f=\phi \let\c=\chi \let\y=\psi
\let\vp=\varphi \let\vep=\varepsilon
      \let\G=\Gamma \let\D=\Delta \let\Th=\Theta \let\L=\Lambda
\let\X=\Xi \let\P=\Pi \let\S=\Sigma \let\U=\Upsilon \let\Y=\Psi
\let\C=\Chi
\let\la=\label \let\ci=\cite \let\re=\ref
\let\se=\section \let\sse=\subsection \let\ssse=\subsubsection
\def\nn{\nonumber} \def\bd{\begin{document}} \def\ed{\end{document}}
\def\ds{\documentstyle} \let\fr=\frac \let\bl=\bigl \let\br=\bigr
\let\Br=\Bigr \let\Bl=\Bigl
\let\bm=\bibitem
\let\na=\nabla
\def\tU{{\widetilde U}}
\let\pa=\partial \let\ov=\overline
\def\ie{{\it i.e.\ }}
\def \be {\begin{equation}}
\def \ee {\end{equation}}
\def\ba{\begin{array}}
\def\ea{\end{array}}
\def\ft#1#2{{\textstyle{{\scriptstyle #1}\over {\scriptstyle #2}}}}
\def\fft#1#2{{#1 \over #2}}
\def\F#1#2{{ F_{#1}^{(#2)} }}
\def\cF#1#2{{ {\cal F}_{#1}^{(#2)} }}

\def\R{{\bf R}}
\def\sst#1{{\scriptscriptstyle #1}}
\def\oneone{\rlap 1\mkern4mu{\rm l}}
\def\e7{E_{7(+7)}}
\def\td{\tilde}
\def\wtd{\widetilde}
\def\im{{\rm i}}
\def\bog{Bogomol'nyi\ }


\def\l {\lambda}
\def\a {\alpha}
\def\ap {\alpha'}
\def\b {\beta}
\def\g {\gamma}
\def\G {\Gamma}
\def\d {\delta}
\def\s {\sigma}
\def\e {\epsilon}
\def\vt {\vartheta}
\def\vp {\varphi}
\def\T {\Theta}

\renewcommand{\O}{\Omega}
\renewcommand{\L}{\Lambda}
\renewcommand{\t}{\theta}


\def \rt {{\rm t}}
\def \ci{\cite}
\def \YY {{\rm Y}}
\def \lra {\leftrightarrow}
\def \const {{|rm const}}

\def\be{\begin{equation}}
\def\ee{\end{equation}}
\def\ba{\begin{eqnarray}}
\def\ea{\end{eqnarray}}

\def\dg{\dagger}
\def\a{\alpha}
\def\b{\beta}
\def\e{\varepsilon}
\def\p{\phi}
\def\ap{\alpha^\prime}
\def\I{{\cal I}}

\def\R{{\bf R}}
\def\Z{{\bf Z}}
\def\C{{\bf C}}
\def\P{{\bf P}}
\def\xb{{\bar X}}
\def\Tr{{\rm  Tr}}
\def\tr{{\rm  tr}}
\def \del{\partial}
\def \a {\alpha}
\def \aa {{\a'}}
\def\g{\gamma}
\def\s{\sigma}
\def\z{\zeta}
\def\zi{\zeta_1}
\def\zii{\zeta_2}
\def\ov{\over}
\def\la{\label}
\def\I{{\cal I}}
\def\J{{\mathcal J}}
\def \ok {{1\ov \k}}
\def\LL{{\mathcal L }}
\def\RR{{\mathcal R }}
\def \jL {{J}}
\def \om {\omega}
\def \cL {{\mathcal L}} \def \cH {{\mathcal H}}
\def\E{{\mathcal E}}
\def\w{\omega}
\def\b{\beta}
\def\l{\lambda}
\def\eps{\epsilon}
\def\vep{\varepsilon}
\def \De {{\mathcal D}}
\def\ttp{\tilde{\tilde{\phi}}}
\def\tp{\tilde{\phi}}
\def\ttvp{\tilde{\tilde{\varphi}}}
\def\tvp{\tilde{\varphi}}
\def\ttpsi{\tilde{\tilde{\psi}}}
\def\tpsi{\tilde{\psi}}
\def \adss{$AdS_5 \times S^5$\ }

\def \r { \rho}
\def \sql {\sqrt{\lambda} }
\def \t {\theta}
\def \p {\phi}
\def \vp {\varphi}
\def \Om {\Omega}
\def \ads {{$AdS_5$}}
\def \ov {\over}
\def \s{\sigma}
\def \pa{\partial}
\def \ta{\tau}
\def \sh {\sinh}
\def \ha {{1 \over 2}}

\def \ttP{\tilde{\tilde{\Phi}}}

\def \la{\label}
\def  \Jt {  {J}_{\rm tot}    }
\def\LM {Lunin-Maldacena }
\def\LMb {$\beta$-deformed }
\def\LMg {$\gamma$-deformed }
\def \k {\kappa}
\def\foot{\footnote}
\def \four{{\textstyle {1\ov 4}}}
 \def \third { \textstyle {1\ov 3}}
\def\det{\hbox{det}}
\def \ci {\cite}

\def \foot {\footnote}
\def \bi{\bibitem}
\def \tr {{\rm tr}}
\def \ha {{1 \over 2}}
\def \tid {\tilde}
\def \vv {{\rm v}}
\def \XX {{\rm X}}
\def \ta {{\tilde \a}}
\def \fo { {1\ov 4}}
\def \ep {\epsilon}
\def \inti {{\int^{2\pi}_0 {d \sigma \ov 2 \pi}}}

\def \el {\ell}
\def \Tr {{\rm Tr}}
\def \P {\Phi}
\def \l  {\lambda}
\def \tl {{\tilde \l}}
\def \bl {{\tilde \l}}
\def \const {{\rm const}}
\def \V {v}

\def \bv {v^*}
\def \vv {{\rm v}}
\def \LL {{\mathcal L}}
\newcommand{\PV}[1]{P_{\!\!_{V_{#1}}}}

\def \bL {\ell}
\def \M {{\mathcal M}}
\def \N {{\mathcal N}}
\def \A {{\mathcal A}}
\def \S {{\rm S}}
\def \vn {\vec n}
\def \tl {\td \l}
\def \td {\tilde}
\def \Prod {\Pi}\def \O {{\mathcal O}}
\def \Q {{\rm  Q}}
\def \D {\Delta}
\def \N {{\mathcal N}}

\def\ve{\varepsilon}
\def\vf{\varphi}
\def\F{\Phi}
\def\wg{\wedge}

\def\ve{\varepsilon}
\def\vf{\varphi}
\def\F{\Phi}
\def\wg{\wedge}

\newcommand{\auth}{AUTHORS}
\def\thb{\bar{\theta}}
\def\Thb{\bar{\Theta}}
\def\barp{\bar{p}}
\def\barq{\bar{q}}
\def\barc{\bar{c}}
\def\bard{\bar{d}}
\def\e{\epsilon}

\def\th{\theta}
\def\Th{\Theta}
\def\vth{\vartheta}
\def\btheta{{\bar\theta}}
\def\ttheta{{{\tilde\theta}}}
\def\bttheta{{{\bar\ttheta}}}
\def\vth{\vartheta}

\def\hg{\hat{\gamma}}

\def\ra{\rightarrow}
\def\N{{\cal N}}
\def\F{{\cal F}}
\def\uM{\underline{M}}
\def\uN{\underline{N}}
\def\uP{\underline{P}}
\def\cc{\circ}
\def\eqv{\equiv}

\def\ni{\noindent}

\def\Ep{E^{{}^{(+)}}}
\def\Em{E^{{}^{(-)}}}

\def\Mp{M^{{}^{(+)}}}
\def\Mm{M^{{}^{(-)}}}

\def\xb{\bar{x}}
\def\xib{\bar{\xi}}
\def\lb{\bar{\l}}
\def\vb{\bar{v}}

\def\an{|n|}
\def\xt{\tilde{x}}
\def\pnr{(p_n^{(r)}}
\def\Xd{\dot{X}}
\def\amn{{\a_{-n}}}
\def\At{\tilde{A}}
\def\Bt{\tilde{B}}
\def\ola{\overleftarrow}
\def\ora{\overrightarrow}
\def\at{\tilde{\a}}
\def\st{\star}
\def\qb{\bar{q}}
\def\qt{\tilde{q}}
\def \w {\omega}

\def \adss {$AdS_5 \times S^5\ $}

\def\ov{\over}
\def \ha { { 1 \over 2}}
\def \kk {{\rm k}}
\def \om {\omega}
\def \Om {\Omega}
\def \X {{\rm  X}}
\def \ww { {\rm w} }
\def \JJ {{\cal I}}
\def \bx {{\bar X}}
\def \del {\pa}
\def \rk {\mbox k}


\newcommand{\bea}{\begin{eqnarray}}
\newcommand{\eea}{\end{eqnarray}}

\newcommand{\beqr}{\begin{displaymath}}
\newcommand{\eeqr}{\end{displaymath}}
\newcommand{\beqa}{\begin{eqnarray}}
\newcommand{\eeqa}{\end{eqnarray}}
\newcommand{\beqar}{\begin{eqnarray*}}
\newcommand{\eeqar}{\end{eqnarray*}}
\renewcommand{\k}{\kappa}

\def\lt {\tilde{\lambda}}
\def\rt {\tilde{r}}
\def\rhot {\tilde{\rho}}
\def \rvac{r_\mt{vac}}
\def \S{{\cal S}}

\newcommand{\rf}[1]{(\ref{#1})}


\def\appendix#1{
  \addtocounter{section}{1}
  \setcounter{equation}{0}
  \renewcommand{\thesection}{\Alph{section}}
  \section*{Appendix \thesection\protect\indent \parbox[t]{11.15cm}
  {#1} }
  \addcontentsline{toc}{section}{Appendix \thesection\ \ \ #1}
  }

\overfullrule=0pt
\parskip=2pt
\parindent=12pt
\headheight=0in \headsep=0in \topmargin=0in
\oddsidemargin=0in
\vspace{ -2cm}
\begin{flushleft}
\hfill hep-th/0503201
\end{flushleft}

\bigskip
\thispagestyle{empty}

\vspace{1cm}
\begin{center}
{\Large\bf Lax Pair for Strings in Lunin-Maldacena Background}

\vspace{0.2cm}

\vspace{.5cm} {Sergey Frolov\footnote{Also at Steklov
Mathematical Institute, Moscow. frolovs@sunyit.edu}}
\vskip 0.3cm

{\em Department of Mathematics/Science,
SUNY IT,\\
P.O. Box 3050, Utica, NY 13504-3050, USA}

\end{center}

 \vspace{0.2cm}
 \begin{abstract}
Recently Lunin and Maldacena used an $SL(3,R)$ transformation of the
$AdS_5\times S^5$ background to generate a supergravity solution
dual to a so-called $\beta$-deformation of $\N =4$ super Yang-Mills
theory. We use a T-duality-shift-T-duality (TsT) transformation to obtain
the $\b$-deformed background for real $\b\equiv\g$, and show that solutions
of string theory equations of motion in this background are in one-to-one
correspondence with those in  $AdS_5\times S^5$ with
twisted boundary conditions imposed on the $U(1)$ isometry fields.
We then apply the TsT transformation to derive a local and periodic Lax pair for
the bosonic part of string theory in the $\g$-deformed background. We also
perform a chain of three consecutive TsT transformations to generate
a three-parameter deformation of $AdS_5\times S^5$. The three-parameter background
is dual to a nonsupersymmetric marginal deformation of $\N=4$ SYM.
Finally, we combine the TsT transformations with $SL(2,R)$ ones to obtain
a $6+2$ parameter deformation of $AdS_5\times S^5$.

\end{abstract}

\newpage
\setcounter{footnote}{0}
\renewcommand{\theequation}{1.\arabic{equation}}
 \setcounter{equation}{0}

\section{Introduction}
In a recent paper \cite{LM}, Lunin and Maldacena have found a supergravity
background which they have conjectured to be
dual to a marginal deformation of $\N =4$ SYM sometimes called
a $\b$ deformation \cite{LEST,BELE,Aharony,DHK,Niarchos,RR,BECH}.

A relative simplicity of the \LM supergravity background and the $\N=1$ superconformal theory
makes the conjectured duality a new promising arena for studying the AdS/CFT correspondence
\cite{MALD,GKPO,WITT}. It is definitely worth to perform tests of the duality by using
methods and ideas developed for the simplest example of the AdS/CFT correspondence
between superstrings on \adss and $\N=4$ SYM. One of such tests has been already performed in
\cite{LM}, where the plane-wave limit of the \LMb background has been analyzed, and it has been
shown that the spectrum of string theory in the pp-wave coincides
with the spectrum of BMN-type operators \cite{BMN} in the \LMb $\N=4$ SYM \cite{Niarchos}.

The BMN operators are dual to heavy semiclassical string states which can be
analyzed by means of semiclassical methods \cite{GKP2,FT1}. It has been found in \cite{FT}
that there is a large class of multi-spin string solutions in \adss whose energies
in a special scaling limit admit an expansion in powers of the t' Hooft coupling $\l$, and,
therefore, can be compared with perturbative anomalous dimensions of dual $\N=4$ SYM operators.

Computation of conformal dimensions of ``long'' $\N=4$ SYM operators became feasible after it was
realized in \cite{MZ} that in the scalar fields sector
the one-loop dilatation operator of $\N=4$ SYM coincides with the
Hamiltonian of an integrable spin chain, and, therefore, one can use the Bethe ansatz
for the chain to analyze the spectrum of the dilatation operator. The integrability
of the complete one-loop dilatation operator was then demonstrated in \cite{BS},
and the higher-loop
integrability of $\N=4$ SYM was discussed in \cite{BKSt,SS,BDS,AFST,Staud}, see \cite{Beis}
for a review and references.

The Bethe ansatz for the dilatation operator allowed to perform comparisons
of field theory and string theory results, first by matching energies of the explicitly
known examples of multi-spin string solutions in \adss with conformal dimensions of
$\N=4$ SYM operators
\cite{BMSZ,AFRT,AS,Upp,Min3,Kop}, and then in general by using the hidden relation of
the integrable
structures of the semiclassical string theory and the spin chain describing long SYM
operators \cite{Kruc,KMMZ,Kruc2,Kruc3,KMMZ2}, see \cite{Ts}
for a review and references.

It is interesting to understand if similar comparisons can be performed for the \LM case.
It is known that the one-loop dilatation operators in several sectors of the \LMb $\N=4$
SYM theory can be identified with Hamiltonians of integrable spin chains \cite{RR,BECH}.
Thus, anomalous dimensions of primary operators in these sectors can be computed by
means of the Bethe ansatz technique.

On the other hand, in the approach of \cite{KMMZ}, the integrability of string sigma model played
a crucial role in the proof of the one- and, especially, two-loop agreement and three-loop
disagreement \cite{SS} between string theory and gauge theory computations.
The Lax representation for superstring theory in \adss \cite{BPR} allows one to
derive a system of singular integral equations called string Bethe equations \cite{KMMZ,KMMZ2}
which describes multi-spin string states.
At one- and two-loops of SYM
perturbation theory the string Bethe equations appear to be equivalent
to the thermodynamic limit of
the spin chain Bethe equations, thus providing a proof of the matching
of multi-spin string energies with conformal dimensions of SYM operators.

Our aim in this paper will be to find a Lax representation for strings in the \LMb background
for real $\b\equiv\g$, which can be used to derive the string Bethe equations.

The $\b$-deformed background was derived
from $AdS_5\times S^5$ by using the $SL(3,R)$
symmetry of type IIB supergravity compactified on a 2-torus.
On the other hand, if the parameter
of the deformation is real,
one can obtain the \LMg  background from $AdS_5\times S^5$ by means of
a T-duality transformation on a $U(1)$ isometry variable $\vp_1$,
a shift of another isometry variable, followed by another T-duality on $\vp_1$
\cite{LM}.\footnote{T-duality
and shifts were used in \cite{RUTS} to derive Melvin-type solutions from flat space.}
We will refer to the chain of these transformations as a TsT-transformation.
Then, the solution with complex $\b$ is obtained by performing  $SL(2,R)_s$ transformations
of the 10-dimensional type IIB supergravity.

In this paper we will discuss the TsT-transformation in detail.
In section 2 we re-derive the metric and the two-form field part of the
\LM background in the real deformation parameter case by
using the TsT-transformation, and show that classical solutions
of string theory equations of motion in their background are in one-to-one
correspondence with those in the $AdS_5\times S^5$ background with
twisted boundary conditions imposed on the $U(1)$ isometry fields.
An interesting property of the twist is that it depends on the
conserved $U(1)$ charges of the model, and, therefore, the space of solutions is
divided into sectors characterized by the charges. There is no twist if
all the three charges are equal, $J_1=J_2=J_3$, and, therefore, all $(J,J,J)$
string solutions can be obtained from usual periodic string solutions in $AdS_5\times S^5$.

In section 3 we apply the TsT transformation to derive a local and periodic Lax pair for
the bosonic part of string theory in the $\g$-deformed background.

In section 4 we discuss the Lax representation for the
simplest reduction of the  string sigma model in
the \LM background to two-spin string states in $R\times S^3_\g$. These states
are dual to operators from the simplest closed $su(2)_\g$ subsector of
the $\g$-deformed  $\N=4$ SYM that consists of operators
made out of two holomorphic scalars.

In section 5 we
perform a chain of  consecutive TsT transformations on each
of the three ``natural'' tori of $S^5$
with different shift parameters $\hg_i$ to generate a nonsupersymmetric deformation of
\adss. If all the parameters $\hg_i$ are equal to each other the deformation reduces to the
\LM one. The three-parameter supergravity background should be dual to a nonsupersymmetric
but marginal deformation of $\N=4$ SYM, and we discuss the scalar fields part of
the deformed YM potential.
Since a general \LM transformation
can be symbolically written as $S\, TsT\, S^{-1}$, where $S$ denotes an $SL(2,R)$
transformation, we also get a 6+2 parameter solution from \adss by performing
three consecutive STsTS transformations on each of the three tori.

Our conclusions are summarized in section 6.

In Appendix A we present the T-duality
transformation in the form used in this paper, and in Appendix B we write down
the $6+2$ parameter deformation of \adss.

\renewcommand{\theequation}{2.\arabic{equation}}
 \setcounter{equation}{0}

\section{TsT-transformation}
In this section we consider string theory sigma model action on \adss, and
derive the metric and the two-form field part of
the \LM background in the case of real $\b\equiv\g$
by using a T-duality on one circle of
$S^5$, a shift of a second angle variable, followed by another
T-duality. It is sufficient to know only the dependence of the string
action on $g_{mn}$ and $B_{mn}$ to use it for the semiclassical analysis of
multi-spin string states.

Since the TsT-transformation involves only
variables of $S^5$ it is sufficient to  consider the $S^5$ part of the string
action that can be written in the form
\bea
\la{a1}
\tilde{\tilde{S}} = -{\sqrt{\lambda}\over 2}\int\, d\tau {d\s\over
2\pi} \left[ \g^{\a\b}\left( \partial_\a r_i\partial_\b r_i +
r_i^2\partial_\a \tilde{\tilde{\phi}}_i\partial_\b \ttp_i\right) +
\Lambda (r_i^2 - 1)\right]\, .
\eea
Here $\sqrt{\l} = R^2/\a'$, $R$
is the radius of $S^5$, $\Lambda$ is a Lagrange multiplier,
$i=1,2,3$, and $\g^{\a\b}\equiv \sqrt{-h}\, h^{\a\b}$, where
$h^{\a\b}$ is a world-sheet metric with Minkowski signature. In
the conformal gauge $\g^{\a\b} = \mbox{diag}(-1,1)$ but we do not
fix the world-sheet metric in this section. The action is
invariant under the $SO(6)$ group, and the three $U(1)$ isometry
transformations are realized as shifts of the angle variables
$\ttp_i$.

To derive the \LMg background it is convenient to make the
following change of variables \cite{LM}
\bea
\la{change1}
\ttp_1=\ttvp_3 - \ttvp_2\, ,\qquad \ttp_2=\ttvp_3+\ttvp_1+\ttvp_2\,
,\qquad \ttp_3=\ttvp_3 - \ttvp_1 \, ,\qquad \ttvp_3\equiv\ttpsi\, .
\eea
In terms of these new angle variables the action (\ref{a1})
takes the form
\bea
\la{a2} \tilde{\tilde{S}} =
-{\sqrt{\lambda}\over 2}\int\, d\tau {d\s\over 2\pi} \left[
\g^{\a\b}\left( \partial_\a r_i\partial_\b r_i + g_{ij}\,
\partial_\a \tilde{\tilde{\vp}}_i\partial_\b \ttvp_j\right) +
\Lambda (r_i^2 - 1)\right]\, ,
\eea
where the metric components $g_{ij}$ are
\bea
\la{metric1} g_{11} = r_2^2 + r_3^2\, ,\qquad
&&g_{22}=r_1^2+r_2^2\, ,\qquad g_{33}=1\, ,\\ \nonumber
g_{12}=r_2^2\, ,\qquad &&g_{13}=r_2^2 - r_3^2\ ,\qquad
g_{23}=r_2^2-r_1^2\, .
\eea
Then we make the T-duality
transformation on the circle parameterized by $\ttvp_1$. By using
the formulas collected in Appendix A, we get the action for the
T-dual theory
\bea
\la{a3} \tilde{S} = -{\sqrt{\lambda}\over
2}\int\, d\tau {d\s\over 2\pi} \left[ \g^{\a\b}\left( \partial_\a
r_i\partial_\b r_i +\tilde{g}_{ij}\partial_\a
\tilde{\vp}_i\partial_\b \tvp_j\right) -
\e^{\a\b}\tilde{b}_{ij}\partial_\a \tilde{\vp}_i\partial_\b \tvp_j
+ \Lambda (r_i^2 - 1)\right]\, .
\eea
Here $\e^{01}=1$, the T-transformed metric
$\tilde{g}$ and the skew-symmetric B-field $\tilde{b}_{ij}$ are
given by
\bea
\nonumber && \tilde{g}_{11} = {1\over r_2^2 +
r_3^2}\ ,\ \ \ \tilde{g}_{22}={r_1^2 r_2^2 + r_1^2 r_3^2 + r_2^2
r_3^2\over r_2^2 +  r_3^2}\ ,\ \ \
\tilde{g}_{33}=1-{(r_2^2-r_3^2)^2\over r_2^2+r_3^2}\ , \ \ \
\tilde{g}_{12}=\tilde{g}_{13}= 0\, ,\\
\nonumber && \tilde{g}_{23}=  {2 r_2^2 r_3^2-r_1^2r_2^2 - r_1^2
r_3^2 \over r_2^2+r_3^2}\ , \ \ \ \tilde{b}_{12} = {r_2^2\over
r_2^2 + r_3^2}\ ,\ \ \ \tilde{b}_{13} ={r_2^2-r_3^2\over r_2^2 +
r_3^2}\ ,\ \ \ \tilde{b}_{23} = 0 \, .
\eea
The T-dual variables
$\tvp_i$ are related to $\ttvp_i$ as follows
\bea
\la{angle1}
 \e^{\a\b}\partial_\b\tvp_1=\g^{\a\b}\partial_\b\ttvp_i\, g_{1i}
&\Leftrightarrow  &\partial_\a\ttvp_1 =
\g_{\a\b}\e^{\b\g}\partial_\g\tvp_1\, \tilde{g}_{11} -
\partial_\a\tvp_i\, \tilde{b}_{1i}\, ,\\
\nonumber \ttvp_2 = \tvp_2\, , \ \ \ \ \ &&\ttvp_3=\tvp_3\ .
\eea
The relations (\ref{angle1}) are satisfied only on-shell, that
means that their consistency conditions  lead to the equations of
motion for $\ttvp_i$ and $\tvp_i$.

Next, we make
the following shift of the angle variable $\tvp_2$
\bea
\la{shift}
\tvp_2 \to \tvp_2 +\hg\, \tvp_1\, ,
\eea
where $\hg$ is any
constant. After the shift the T-transformed metric $\tilde{g}$
acquires the following form, $\tilde{g}_{ij}\to \tilde{G}_{ij}$:
\bea
\nonumber
&& \tilde{G}_{11} = \tilde{g}_{11} + \hg^2\,
\tilde{g}_{22} = {G^{-1}\over r_2^2+r_3^2}\, ,\ \ \
G^{-1}=1+\hg^2\,(r_1^2r_2^2+r_1^2r_3^2+r_2^2r_3^2)\, ,\ \ \ \ \\
\nonumber &&\tilde{G}_{22} = \tilde{g}_{22}\ ,\ \ \ \tilde{G}_{33}
= \tilde{g}_{33}\ ,\ \ \ \  \tilde{G}_{12} = \hg\, \tilde{g}_{22}\
,\ \ \ \  \tilde{G}_{13} = \hg\, \tilde{g}_{23}\ ,\ \ \ \
 \tilde{G}_{23} = \tilde{g}_{23}\, ,
\eea
and the eq.(\ref{angle1}) transforms into
\bea
\la{angle2}
\partial_\a\ttvp_1 =
\g_{\a\b}\e^{\b\g}\partial_\g\tvp_1\, \tilde{g}_{11} -
\partial_\a\tvp_i\, \tilde{b}_{1i} - \hg\, \partial_\a\tvp_1\,
\tilde{b}_{12}\, .
\eea
The final step is to make again the
T-duality transformation on the circle parameterized by $\tvp_1$.
This leads to the string action on the \LMg background
\bea
\la{a4}
S = -{\sqrt{\lambda}\over 2}\int\, d\tau {d\s\over
2\pi} \left[ \g^{\a\b}\left( \partial_\a r_i\partial_\b r_i
+G_{ij}\partial_\a \vp_i\partial_\b \vp_j\right) -
\e^{\a\b}B_{ij}\, \partial_\a \vp_i\partial_\b \vp_j + \Lambda
(r_i^2 - 1)\right]\, .
\eea
The variables $\tvp_i$ are related to
the T-dual variables $\vp_i$ as follows
\bea
\la{angle3}
 \e^{\a\b}\partial_\b\vp_1=\g^{\a\b}\partial_\b\tvp_i\, \tilde{G}_{1i}
 -\e^{\a\b}\partial_\b\tvp_i\, \tilde{b}_{1i}
&\Leftrightarrow  &\partial_\a\tvp_1 =
\g_{\a\b}\e^{\b\g}\partial_\g\vp_i\, G_{1i} -
\partial_\a\vp_i\, B_{1i}\, ,\\
\nonumber \tvp_2 = \vp_2\ , \ \ \ \ \ &&\tvp_3=\vp_3\, .
\eea
The eqs.(\ref{angle1}), (\ref{angle2}) and (\ref{angle3}) allow us to
determine the following relations between the angle variables
$\ttvp_i$ and the TsT-transformed variables $\vp_i$:
\bea
\la{angle4}
&&\partial_\a\ttvp_1
=\left(\tilde{g}_{11}G_{1i}+\hg\, \tilde{b}_{12}B_{1i}
-\tilde{b}_{1i}\right)\partial_\a\vp_i -\left( \hg\,
\tilde{b}_{12}G_{1i} + \tilde{g}_{11}B_{1i}\right)
\g_{\a\b}\e^{\b\g}\partial_\g\vp_i\, ,\\
\nonumber &&\partial_\a\ttvp_2 =\partial_\a\vp_2 -\hg\,
B_{1i}\partial_\a\vp_i\
+\hg\,G_{1i}\g_{\a\b}\e^{\b\g}\partial_\g\vp_i\ ,\\ \nonumber
&&\partial_\a\ttvp_3 =\partial_\a\vp_3\, .
\eea

The \LMg metric in (\ref{a4}) is given by
\bea
\nonumber
G_{ij} = G\, g_{ij}\ , \ \mbox{if both } i,j\neq
3\ ;\qquad  G_{33} = G\,g_{33} + 9\,\hg^2\, G\,
r_1^2r_2^2r_3^2\ .
\eea It is easy to see from this form of the
metric that in terms of the angle variables $\phi_i$,
eq.(\ref{change1}), the action takes the following simple form
\bea
\nonumber
S = -{\sqrt{\lambda}\over 2}\int\, d\tau {d\s\over
2\pi} \Big[ \g^{\a\b}\left( \partial_\a r_i\partial_\b r_i + G\,
r_i^2\partial_\a \phi_i\partial_\b \phi_i+ \hg^2\, G\,
r_1^2r_2^2r_3^2\big(\sum_i\partial_\a
\phi_i\big)\big(\sum_j\partial_\b \phi_j\big)\right) &&\\
\la{a5} - 2\, \hg\, G\, \e^{\a\b}\left(r_1^2r_2^2\partial_\a
\phi_1\partial_\b \phi_2 + r_2^2r_3^2\partial_\a \phi_2\partial_\b
\phi_3 + r_3^2r_1^2\partial_\a \phi_3\partial_\b \phi_1 \right) +
\Lambda (r_i^2 - 1)\Big]\, . \ \ \ \ \ \  &&
\eea
It is in this form
the gravity background was written in \cite{LM}.

The relations (\ref{angle4}) also acquire a very nice and symmetric
form being rewritten in terms of $\p_i$:
\bea
\la{angle4b}
&&\partial_\a\ttp_1 =G\, \left(\partial_\a\p_1\, +\, \hg^2\, r_2^2r_3^2 \sum_i\partial_\a
\phi_i \,  -\, \hg\,
\g_{\a\b}\e^{\b\g}\left( r_2^2\partial_\g\p_2\, -\, r_3^2\partial_\g\p_3 \right)\right)
\, ,\\ \nonumber
&&\partial_\a\ttp_2 =G\, \left(\partial_\a\p_2\, +\, \hg^2\, r_1^2r_3^2 \sum_i\partial_\a
\phi_i \,  -\, \hg\,
\g_{\a\b}\e^{\b\g}\left( r_3^2\partial_\g\p_3\, -\, r_1^2\partial_\g\p_1 \right)\right)
\, ,\\ \nonumber
&&\partial_\a\ttp_3 =G\, \left(\partial_\a\p_3\, +\, \hg^2\, r_1^2r_2^2 \sum_i\partial_\a
\phi_i \,  -\, \hg\,
\g_{\a\b}\e^{\b\g}\left( r_1^2\partial_\g\p_1\, -\, r_2^2\partial_\g\p_2 \right)\right)
\, .
\eea
In next section we  use these
relations to find the Lax pair for string theory  on the
\LMg background.

To clarify the meaning of the relations (\ref{angle4b}) we compute the conserved
$U(1)$ isometry currents for the string theories on the backgrounds under consideration
\bea
\la{currents}
&&\tilde{\tilde{J}}^\a_i = -\sqrt{\l}\,r_i^2\, \g^{\a\b}\, \partial_\b\ttp_i\, ,\\ \nonumber
&&J_1^\a =-\sqrt{\l}\, r_1^2\, \g^{\a\b}\, G\,\left(\partial_\b\p_1\,
+\, \hg^2\, r_2^2r_3^2 \sum_i\partial_\b
\phi_i \,  -\, \hg\,
\g_{\b\r}\e^{\r\g}\left( r_2^2\partial_\g\p_2\, -\, r_3^2\partial_\g\p_3 \right)\right)
\, ,\\ \nonumber
&&J_2^\a =-\sqrt{\l}\, r_2^2\, \g^{\a\b}\, G\, \left(\partial_\b\p_2\,
+\, \hg^2\, r_1^2r_3^2 \sum_i\partial_\b
\phi_i \,  -\, \hg\,
\g_{\b\r}\e^{\r\g}\left( r_3^2\partial_\g\p_3\, -\, r_1^2\partial_\g\p_1 \right)\right)
\, ,\\ \nonumber
&&J_3^\a =-\sqrt{\l}\, r_3^2\, \g^{\a\b}\, G\, \left(\partial_\b\p_3\,
+\, \hg^2\, r_1^2r_2^2 \sum_i\partial_\b
\phi_i \,  -\, \hg\,
\g_{\b\r}\e^{\r\g}\left( r_1^2\partial_\g\p_1\, -\, r_2^2\partial_\g\p_2 \right)\right)
\, .
\eea
Comparing the relations (\ref{angle4b}) with the expressions for the currents (\ref{currents}),
we see that (\ref{angle4b}) is just a statement that the $U(1)$ currents of strings
on $S^5$ are equal to
those on the \LMg background:
\bea
\la{currents2}
\tilde{\tilde{J}}^\a_i\, =\, J^\a_i\ .
\eea
We expect that the same relations hold for any two backgrounds related by a TsT transformation.

To get more insight into the relations (\ref{currents2}) let
us consider them for the time and space
components. The time component $J_i^0$ of the current $J^\a_i$ is just the momentum
conjugated to the angle variable $\p_i$, and the charge $J_i$ is equal to the integral
of $J^0_i$ over $\s$:
\bea
\la{momentum}
p_i \, =\,  J_i^0\, ,\quad J_i\, =\, \int\, {d\s\ov 2\pi}\, J_i^0\, ,\qquad
\tilde{\tilde{p}}_i\, =\, \tilde{\tilde{J}}_i^0\, ,
\quad \tilde{\tilde{J}}_i\, =\, \int\, {d\s\ov 2\pi}\, \tilde{\tilde{J}}_i^0\, .
\eea
Thus, the time component of the relations (\ref{currents2}) says that the momenta
do not change under the TsT transformation:
\bea\la{mome}
\tilde{\tilde{p}}_i \=  p_i\, .
\eea
To analyze the space component
of the relations (\ref{currents2}), $\tilde{\tilde{J}}^1_i\, =\, J^1_i$,
we express the time derivatives
$\pa_0\phi_i \equiv \dot{\phi}_i$ and $\dot{\ttp}_i$ through momenta
$p_i$ and $\tilde{\tilde{p}}_i$, and substitute them into (\ref{currents2}). Then
the relations $\tilde{\tilde{J}}^1_i\, =\, J^1_i$ take the following simple form
\bea
\la{dphi}
&&\ttp_1' \= \p_1'\+ \g\,(p_2 \- p_3)\, ,\\ \nonumber
&&\ttp_2' \= \p_2'\+ \g\,(p_3 \- p_1)\, ,\\ \nonumber
&&\ttp_3' \= \p_3'\+ \g\,(p_1 \- p_2)\, ,
\eea
where
$$\g\= {\hg\ov\sqrt{\l}}$$
is the deformation parameter that appears in field theory. Taking into account that $\p_i$
are angle variables, and integrating (\ref{dphi}) over $\s$, we get the following
twisted boundary conditions for the $U(1)$ variables $\ttp_i$ of the \adss theory:
\bea
\la{dphi2}
&&\p_i(2\pi) \- \p_i(0) \= 2\pi\, n_i\, ,\quad n_i \mbox{ are integer winding numbers}\, , \\
\nonumber
&&\ttp_1(2\pi) \- \ttp_1(0) \= 2\pi \left(n_1\+ \g\,(J_2 \- J_3)\right)\, ,\\ \nonumber
&&\ttp_2(2\pi) \- \ttp_2(0) \= 2\pi \left(n_2\+ \g\,(J_3 \- J_1)\right)\, ,\\ \nonumber
&&\ttp_3(2\pi) \- \ttp_3(0) \= 2\pi \left(n_3\+ \g\,(J_1 \- J_2)\right)\, .
\eea
Since the equations of motion for the $U(1)$ variables have the form
$\pa_\a J^\a_i \= 0\, ,\pa_\a \tilde{\tilde{J}}^\a_i \= 0$, the relations
(\ref{currents2}) imply that if $\p_i$ solve equations of motion
for strings in the \LMg background then $\ttp_i$ solve those in \adss with
the twisted boundary conditions (\re{dphi2}) imposed on $\ttp_i$, and vice versa.
It is not difficult to check that the Virasoro constraints for both models map to each other
under the TsT transformation, and, therefore, the energy of such a twisted string in
\adss is equal to the energy of the corresponding string in the \LMg background. The equivalence
between closed strings in the \LMg background and twisted strings in \adss provides an efficient
way of finding multi-spin strings on the \LMg background by using the known results
for the \adss case \cite{FT,AFRT}. In particular, many circular string solutions of
\cite{FT,AFRT}
formally satisfy string equations for noninteger values of winding numbers $n_i$, and, therefore,
the corresponding solutions of string equations for the \LMg background can be obtained
just by shifting the winding numbers of these circular strings
by the twists (\ref{dphi2}). We also see that if
$J_1=J_2=J_3$ then all the twists vanish, and, therefore,
any solution with equal charges $J_1=J_2=J_3$  in the \LMg model
can be obtained from a periodic solution in \adss.\footnote{It is
interesting that as was found in \cite{Min3} the
energy of the 3-spin circular string \cite{FT} with $J_1=J_2=J_3$ matches
the conformal dimension of the dual CFT operator up to the 3-loop order. Generally, the agreement
persists only up to the 2-loop order \cite{SS}.}
Therefore, according to our discussion above,
the energies of that string states in the \LMg model and in \adss are equal to
each other.
Note, however, that the solutions are different because given a solution in \adss
one should still integrate (\ref{dphi}) to find a solution in the \LMg model.

For completeness we rewrite the relations (\ref{dphi}) in terms of the angle variables
$\vp_i$ used in \cite{LM}. We get
\bea
\la{dvphi}
\ttvp_1' \= \vp_1'\+ \g\,\pi_2\, ,\quad
\ttvp_2' \= \vp_2'\- \g\,\pi_1\, ,\quad
\ttvp_3' \= \vp_3'\, .
\eea
Here $\pi_i$ are momenta conjugated to $\vp_i$. They are related to $p_i$ as follows
\bea\nonumber
\pi_1\= p_2\- p_3\, ,\quad \pi_2\= p_2\- p_1\, ,\quad
\pi_3\= p_1\+ p_2\+ p_3\, .
\eea
The twisted boundary conditions for the $U(1)$ variables $\ttvp_i$ of the \adss theory, therefore,
take the form:
\bea
\la{dvphi2}
&&\vp_i(2\pi) \- \vp_i(0) \= 2\pi\, m_i\, ,\\
\nonumber
&&\ttvp_1(2\pi) \- \ttvp_1(0) \= 2\pi \left(m_1\+ \g\,(J_2 \- J_1)\right)\, ,\\ \nonumber
&&\ttvp_2(2\pi) \- \ttvp_2(0) \= 2\pi \left(m_2\+ \g\,(J_3 \- J_2)\right)\, ,\\ \nonumber
&&\ttvp_3(2\pi) \- \ttvp_3(0) \= 2\pi\, m_3\, .
\eea

\renewcommand{\theequation}{3.\arabic{equation}}
 \setcounter{equation}{0}

\section{The Lax pair}
In this section we use the known Lax representation for string
theory on $AdS_5\times S^5$, and the relations (\ref{angle4b}) to
find a local Lax pair for string theory  in the \LMg
background. Since the $AdS_5$ part of the string action is not
modified by the TsT-transformation we can restrict our attention
to the sigma model on $S^5$.

A convenient parametrization of $S^5$ is provided by  unitary
skew-symmetric $SU(4)$ matrices of the form (see, e.g. the second paper of \cite{AFRT}):
\bea
\la{matr}
g\=\left(\begin{array}{cccc}
0 & X_3 & X_1 & X_2 \\
-X_3 & 0 & X_2^* & -X_1^* \\
-X_1 & -X_2^* & 0 & X_3^* \\
-X_2 & X_1^* & -X_3^* & 0
\end{array}
\right) \, , \qquad \det g \= (|X_1|^2+|X_2|^2+|X_3|^2)^2=1 \, .
\eea
The equations of motion for the sigma model on $S^5$ follow
from the usual action for the principal chiral field:
\bea
\nonumber {\rm S}\=\int {\rm d}\tau{\rm d}\sigma
\gamma^{\a\b}{\rm Tr}\Big(g^{-1}\, \pa_{\a}g\,g^{-1}\, \pa_{\b}g \Big) \, ,
\eea
and can be written in the form
\bea
\label{eom}
\pa_{\a}(\gamma^{\a\b}R_{\b})\= 0\, ,
\eea
where we introduce the right current
$$
R_{\a}\=g^{-1}\,\pa_{\a}g\,  .
$$
The equations of motion (\ref{eom}) are equivalent to the zero curvature condition
\cite{lup,ZM,FR,FTa}
\bea
\label{zc}
[D_{\a}\, ,\, D_{\b}]\= 0 \, ,
\eea
where the Lax operator depending on a
spectral parameter $x$ is defined as
\bea
\label{Lax}
D_{\a}\=\pa_{\a}\-\frac{R^{+}_{\a}}{2(x-1)}\+\frac{R^{-}_{\a}}{2(x+1)}\equiv
\pa_{\a}\- {\mathcal A}_{\a}(x) \, . \eea
Here the self- and
anti-self dual projections of $R_\a$ are given by
\bea
\label{proj} R^{\pm}_{\a}\=(P^{\pm})_{\a}^{~\b}R_{\b},~~~~~~~~
(P^{\pm})_{\a}^{~\b}\=\delta_{\a}^{~\b}\mp
\gamma_{\a\rho}\epsilon^{\rho\b} \, .
\eea
The same equations of motion also follow from another Lax operator defined with the help
of the left current $L_{\a}\=\pa_{\a}g\, g^{-1}$
\bea
\label{Lax_left}
D_{\a}^L\=\pa_{\a}\-\frac{L^{+}_{\a}}{2(1-x)}\-\frac{L^{-}_{\a}}{2(1+x)}\equiv
\pa_{\a}\- {\mathcal A}_{\a}^L(x) \, .
\eea
The two Lax operators are related by the following gauge transformation
\bea
\la{RL}
g^{-1}D_{\a}^Lg\=\pa_{\a}\-\frac{R^{+}_{\a}}{2({1\over x}-1)}\+
\frac{R^{-}_{\a}}{2({1\over x}+1)}\equiv
\pa_{\a}\- {\mathcal A}_{\a}({1\over x}) \, . \eea

The Lax operator (\ref{Lax}) for the sigma model on $S^5$
cannot be used to derive a local Lax operator for the sigma model on the \LMg background,
because it is not invariant under the $U(1)$ isometry transformations, and, therefore, has
an explicit dependence on the angle variables $\ttp_i$. The dependence can be easily found
if one notices that the matrix $g$ (\ref{matr})
can be represented in the following factorized form
\bea
\la{factform}
g(r_i,\ttp_i)\=M(\ttp_i)\, \hat{g}(r_i)\, M(\ttp_i)\, , \quad X_i\= r_i\, e^{i\,\ttp_i}\, ,
\eea
where
\bea
\nonumber
M(\ttp_i)\=e^{\tilde{\tilde{\Phi}}}\, ,
\quad \ttP \= {i\over 2}\left(
\begin{array}{cccc}
\ttp_1 + \ttp_2 + \ttp_3&0&0&0\\
0&-\ttp_1-\ttp_2+\ttp_3&0&0\\
0&0&\ttp_1-\ttp_2-\ttp_3&0\\
0&0&0&-\ttp_1+\ttp_2-\ttp_3
\end{array}\right)\, ,&&\\ \la{Mg}
\hat{g}(r_i) = \left(\begin{array}{cccc}
0 & r_3 & r_1 & r_2 \\
-r_3 & 0 & r_2 & -r_1 \\
-r_1 & -r_2 & 0 & r_3 \\
-r_2 & r_1 & -r_3 & 0
\end{array}
\right) \, , \qquad \hat{g}^{-1} = -\hat{g} \, .\ \ \ \ \ \ \ \ \ \ \ \ \ \ \ \ \ \ \ \ \ \
\ \ \ \ \ \ \ \ \ &&
\eea
Using this representation, we get
\bea
\nonumber
R_\a(r_i,\ttp_i)\, =\, M^{-1}\, \hat{R}_\a(r_i,\partial\ttp_i)\, M\, ,
\eea
where
\bea
\hat{R}_\a(r_i,\partial\ttp_i)\, =\,
\hat{g}^{-1}\pa_{\a}\hat{g}\+ \hat{g}^{-1}\,\pa_{\a}\ttP\,\hat{g}\+
\pa_{\a}\ttP \,=\-\hat{g} \pa_{\a}\hat{g}\- \hat{g}\,\pa_{\a}\ttP\,\hat{g}\+
\pa_{\a}\ttP\, .
\eea
It is clear  that the dependence of the Lax connection ${\mathcal R}_{\a}$ on the matrix
$M$ can be gauged away\footnote{The gauge transformation is similar to the one used in \cite{AF}
to derive a local and periodic Lax connection for the Hamiltonian of strings on \adss.}
\bea
\la{gaugetr}
&&D_\a\to M\, D_\a\, M^{-1} \, = \, \pa_\a \, -\, \RR_\a\, ,\\
\nonumber
&&
\RR_{\a}\= M\,{\mathcal
A}_{\a}\,M^{-1}\, -\, M\,\pa_{\a}M^{-1}\, =\, \hat{{\mathcal
A}}_{\a}\, +\, \pa_{\a}\ttP\, .
\eea
The gauged transformed Lax connection $\RR_\a$ is, obviously, flat, and is invariant under
the $U(1)$ isometries, since it depends on $\ttp_i$ only through $\pa_\a\ttp_i$.
Now, to find the Lax representation for the sigma model on the \LMg background
all one needs to do is to express the angle variables, $\pa_\a\ttp_i$, of $S^5$
in terms of the angle variables, $\pa_\a\p_i$, of the \LMg background by using
the explicit relations (\ref{angle4b}). Since the TsT-transformation maps the equations
of motion for the sigma model on $S^5$ to those on the \LM background, the zero-curvature
condition for the new Lax connection is equivalent to the equations of motion
for the sigma model on the \LMg background.

The resulting expression for the Lax connection $\RR_\a$ is rather complicated,
and we refrain from presenting their explicit form. It is interesting, however, to understand
the structure of the Lax component $\RR_1$, because the monodromy matrix $T(x)$ is defined as
its path-ordered exponential. An important property of $\RR_1$ is that it does not
depend on the world-sheet metric $\g^{\a\b}$ if one expresses time derivatives of the fields
$\p_i$ and $r_i$ through their conjugated momenta. It follows from the fact
that
\bea\la{A1pm}
\hat{R}_{1}^\pm \= \hat{R}_{1} \mp \g_{1\r}\e^{\r\b}\hat{R}_{\b}
\= \hat{R}_{1} \mp \g^{0\b}\hat{R}_{\b}\= \hat{R}_{1} \pm \hat{P}
\, ,
\eea
where $\hat{P} \=-\g^{0\b}\hat{R}_{\b}$ is a matrix-valued momentum.
The diagonal and off-diagonal parts of $\hat{P}$ determine the momenta of
$\p_i$ and $r_i$, respectively. The formula (\ref{A1pm}) allows us to
present the Lax component $\RR_1$ of (\ref{gaugetr})
in the form depending only on the coordinates and their
momenta:
\bea
\la{LaxH}
\RR_{1}\, =\,  \pa_{1}\ttP \+ {\hat{R}_1\ov x^2-1} \+  {x\, \hat{P}\ov x^2-1} \, .
\eea
According to (\ref{mome}) and (\ref{dphi}), the momenta are invariant under
the TsT transformation and the coordinates $\ttP$ are just shifted by the momenta.
The formula (\ref{LaxH}) can be used to determine the asymptotic behavior of the monodromy matrix
around $x=\pm 1$. The result in fact coincides with the one for strings in $R \times S^5$ because
all transformations we've done to derive (\ref{LaxH}) do not change the singular
terms in the Lax connection at
$x = \pm 1$, and, therefore, the asymptotic behavior.
A potential problem with this form of $\RR_{1}$ is that it does not vanish
at large values of the spectral parameter $x$, and that may make more difficult to
study the large $x$ asymptotic properties of the monodromy matrix. To study the asymptotics
it is easier to make an inverse gauge transformation, and use a nonlocal and nonperiodic
Lax connection (\ref{Lax}) with the
field $g$ depending on $\ttp_i$ which satisfy the twisted boundary conditions (\ref{dphi}).
Note, however, that the monodromy matrix is not similar to the path-ordered exponential
of the Lax connection (\ref{Lax}) because the inverse gauge transformation is not periodic.
It would be interesting to analyze the properties of the monodromy matrix and derive
the string Bethe equations for the \LM model analogous to those derived for strings
in $AdS_5\times S^5$ in \cite{KMMZ,KMMZ2,AF}.

\renewcommand{\theequation}{4.\arabic{equation}}
 \setcounter{equation}{0}

\section{$su(2)_\g$ subsector}

The simplest closed subsector of the $\g$-deformed  $\N=4$ SYM consists of operators
made out of two holomorphic scalars. These operators are dual to two-spin string states
in $R\times S^3_\g$ which is a consistent reduction of the string sigma model on
the \LM background. We will refer to this subsector as the $su(2)_\g$ subsector.

The $su(2)_\g$ subsector of the \LMg model is obtained by setting $r_3=0$.
It is easy to see that this reduction is compatible with equations
of motion. The action (\ref{a5}) then simplifies drastically
\bea
\la{a6}
S = -{\sqrt{\lambda}\over 2}\int\, d\tau {d\s\over 2\pi}
\Big[ \g^{\a\b}\left( \partial_\a r_i\partial_\b r_i + G\,
r_i^2\partial_\a \phi_i\partial_\b \phi_i\right)  -
2\e^{\a\b}\hg\, G\, r_1^2r_2^2\, \partial_\a
\phi_1\partial_\b\phi_2 + \Lambda (r_i^2 - 1)\Big],
\eea
where $i=1,2$.

This action can be obtained from the string action (\ref{a1}) on
$S^3$ by means of the same TsT-transformation. The only difference
and simplification is that one does not need to make any change of
the angle variables $\ttp_i$ ($i=1,2$). One should just make the
T-duality on $\ttp_1$, the same shift (\ref{shift}), and again the
T-duality on $\tp_1$. The resulting relations between $\ttp_i$ and
$\p_i$ acquire the following simple form
\bea
\la{angle5}
&&\partial_\a\ttp_1 =G\, \left(\partial_\a\p_1 -\hg\,r_2^2
\g_{\a\b}\e^{\b\g}\partial_\g\p_2\right) \ ,\\ \nonumber
&&\partial_\a\ttp_2 =G\, \left(\partial_\a\p_2 +\hg\,r_1^2
\g_{\a\b}\e^{\b\g}\partial_\g\p_1\right)\ ,\\
\nonumber&&G^{-1}=1+\hg^2\, r_1^2r_2^2 .
\eea
We will use these
relations to obtain a local Lax representation for the equations of
motion describing the $su(2)_\g$ subsector.

In the $su(2)_\g$ subsector formulas (\ref{dphi})  reduce
to:
\bea
\la{dphisu2}
\ttp_1' \= \p_1'\+ \g\,p_2\, ,\qquad
\ttp_2' \= \p_2'\- \g\, p_1\, ,
\eea
and the twisted boundary conditions (\ref{dphi2}) take the form
\bea
\la{dphi2su2}
&&\p_i(2\pi) \- \p_i(0) \= 2\pi\, n_i\, ,\quad n_i \mbox{ are integer winding numbers}\, , \\
\nonumber
&&\ttp_1(2\pi) \- \ttp_1(0) \= 2\pi \left(n_1\+ \g\,J_2\right)\, ,\qquad
\ttp_2(2\pi) \- \ttp_2(0) \= 2\pi \left(n_2\- \g\,J_1\right)\, .
\eea
This shift of the winding numbers $n_i$ is consistent with the spectrum
of strings in the \LM background in the pp-wave limit \cite{LM}.

In the $su(2)_\g$ subsector the Lax connection can be written in terms of $2\times 2$ matrices.
We parameterize  $S^3$  by  unitary
$SU(2)$ matrices of the form:
\bea
\la{matrsu2}
g\=\left(\begin{array}{cc}
X_1 & X_2 \\
-X_2^* & X_1^*
\end{array}
\right) \, , \qquad \det g \= |X_1|^2+|X_2|^2=1 \, .
\eea
The Lax operator  for the sigma model on $S^3$ has the same form (\ref{Lax}). It also has
an explicit dependence on the angle variables $\ttp_i$. Representing
the matrix $g$ (\ref{matrsu2})
in the factorized form
\bea
\la{factform}
g(r_i,\ttp_i)\=\tilde{M}(\ttp_i)\,\hat{g}(r_i)\, M(\ttp_i)\, ,
\quad X_i\= r_i\, e^{i\,\ttp_i}\, ,
\eea
where
\bea
\nonumber
&&M\=
\left(\begin{array}{cc}
0 & e^{{i\ov 2}\ttp_2} \\
e^{-{i\ov 2}\ttp_2} & 0
\end{array}
\right)\left(\begin{array}{cc}
e^{{i\ov 2}\ttp_1} & 0 \\
0 & e^{-{i\ov 2}\ttp_1}
\end{array}
\right)\=\left(\begin{array}{cc}
0 & e^{{i\ov 2}(\ttp_2 -\ttp_1)} \\
e^{-{i\ov 2}(\ttp_2-\ttp_1)} & 0
\end{array}
\right)\, ,\\
\nonumber
&&\tilde{M}\=\left(\begin{array}{cc}
e^{{i\ov 2}\ttp_1} & 0 \\
0 & e^{-{i\ov 2}\ttp_1}
\end{array}
\right)
\left(\begin{array}{cc}
0 & e^{{i\ov 2}\ttp_2} \\
e^{-{i\ov 2}\ttp_2} & 0
\end{array}
\right)\=\left(\begin{array}{cc}
0 & e^{{i\ov 2}(\ttp_1 +\ttp_2)} \\
e^{-{i\ov 2}(\ttp_1+\ttp_2)} & 0
\end{array}
\right)\, ,\\
\la{Mgsu2}
&&\hat{g} = \left(\begin{array}{cc}
r_1 & r_2 \\
-r_2 & r_1
\end{array}
\right) \, ,\quad g^{-1}\= g^T\, ,\quad M^{-1}\= M\, ,\quad \tilde{M}^{-1} = \tilde{M}\, ,
\eea
we get
\bea
\nonumber
R_\a(r_i,\ttp_i)\, =\, M^{-1}\, \hat{R}_\a(r_i,\partial\ttp_i)\, M\, ,
\eea
where
\bea
\hat{R}_\a\, =\,
{i\ov 2}(\pa_{\a}\ttp_2 -\pa_\a\ttp_1)\, \s_3\, +\,
\hat{g}^{-1}\pa_{\a}\hat{g}\,  -
{i\ov 2}(\pa_{\a}\ttp_1 +\pa_\a\ttp_2)\,\hat{g}^{-1}\,\s_3\,\hat{g}\,  .
\eea
Gauging away the dependence of the Lax connection ${\mathcal A}_{\a}$ on the matrix, we obtain
the Lax connection that depends only on $\pa_\a\ttp_i$
\bea
\la{gaugetrsu2}
&&D_\a\to M\, D_\a\, M^{-1} \, = \, \pa_\a \, -\, \RR_\a\, ,\\
\nonumber
&&
\RR_{\a}\= M\,{\mathcal
A}_{\a}\,M^{-1}\, -\, M\,\pa_{\a}M^{-1}\, =\, \hat{{\mathcal
A}}_{\a}\, +\, {i\ov 2}(\pa_{\a}\ttp_2 -\pa_\a\ttp_1)\, \s_3\, .
\eea
The local Lax representation for the $su(2)_\g$ subsector
is now obtained by expressing the angle variables, $\pa_\a\ttp_i$, of $S^3$
in terms of the angle variables, $\pa_\a\p_i$, of the $\g$-deformed background by using
the explicit relations (\ref{angle5}).

It is useful to express the Lax component $\RR_1$ in terms of the coordinates
and their conjugated momenta. Introducing the parametrization
$$r_1 \= \cos\t\, ,\quad r_2 \= \sin\t\, ,$$
and
the re-scaled momenta (see (\ref{currents}))
\bea\nonumber
\bp_i \= - \g^{0\a}\pa_\a\ttp_i\, ,\quad \bp_\t \= - \g^{0\a}\pa_\a\t\, ,
\eea
we get the following expressions for the Lax component $\RR_1$ of the \LMg model
\bea
\RR_1 \= i\, \s_3\, \RR_1^{(3)} \+i\, \s_1\, \RR_1^{(1)} \+i\, \s_2\, \RR_1^{(2)} \, ,
\eea
where
\bea\nonumber
&&\RR_1^{(3)}\= {\ttp_2'\- \ttp_1'\ov 2}
\+ {r_2^2\ttp_2'\- r_1^2\ttp_1'\ov x^2-1}
\+ {x\,\left(\bp_2-\bp_1\right)\ov x^2-1}\, ,\\ \nonumber
&&\RR_1^{(1)}\=
-{\, r_1\, r_2\, \left(\ttp_1' + \ttp_2'\right)\ov x^2-1} -
{x\,  \left(r_2^{2}\,\bp_1 + r_1^{2}\,\bp_2\right)\ov (x^2-1)r_1\, r_2}\, ,\\ \nonumber
&&\RR_1^{(2)}\= {x\,\bp_\t\+ \t'\ov x^2-1}\, ,
\eea
and $\ttp_i'$ are expressed through $\p_i'$ by using (\ref{dphisu2}).

The Lax representation (\ref{gaugetrsu2})
for strings in $R\times S^3_\g$ has been recently used to derive the
string Bethe equations \cite{FRT}. It has been shown that these equations coincide with
the one- and two-loop Bethe equations for the spin chain describing the operators
from the holomorphic $su(2)_\g$ subsector \cite{RR,BECH}.
This shows that, just as it was for $\N=4$ SYM, in the
$\g$-deformed case there is a perfect match between string theory and field theory
results at least for the simplest $su(2)_\g$ subsector.

\renewcommand{\theequation}{5.\arabic{equation}}
 \setcounter{equation}{0}

\section{Multi-parameter deformations of $AdS_5\times S^5$}

In this section we use a chain of  TsT transformations to generate
a three-parameter deformation of the $AdS_5\times S^5$ supergravity background.
In the case when all the parameters are equal to each other the deformed background reduces
to the one-parameter Lunin-Maldacena background  we discussed in the previous
sections. At the end of this section we present a 4+2 parameter generalization of
the complex $\b$ \LM solution.

We saw in section 2 that to obtain the \LMg  supergravity background
from $AdS_5\times S^5$ by using a TsT transformation
we had to choose a very special torus in $S^5$.
This choice is related to supersymmetry of the \LM background but in general one may be
interested in studying nonsupersymmetric deformations too. In that case, the choice of the torus
looks rather superficial. On the other hand, in the parametrization
of $S^5$ we use in (\ref{a1}) there are three natural tori: $(\p_1, \p_2)$, $(\p_2,\p_3)$
and $(\p_3,\p_1)$. A TsT transformation applied to any of these tori produces a very simple
one-parameter deformation of \adss similar to the $su(2)_\g$ subsector
of the \LMg background we discussed in section 4.
One may ask how one could get the \LMg background by
using TsT transformations on the three tori. The answer appears to be very simple.
One should just perform a chain of three consecutive TsT transformations on each
of the three tori with the same shift parameter $\hg$. If we allow the TsT transformations
to have different shift parameters $\hg_i$ we get a nonsupersymmetric deformation of
\adss. The three-parameter supergravity background should be dual to a nonsupersymmetric
but marginal deformation of $\N=4$ SYM.

Since the details of the derivation are very similar to the case of the \LMg
background we present here only the final results. We apply the first TsT transformation
with T-duality acting on
the first angle $\p_1$ and the shift parameter equal
to ${\hg_3}$ to the torus $(\p_1, \p_2)$, then the second TsT transformation with
the shift parameter equal to ${\hg_1}$ to the torus $(\p_2, \p_3)$, and finally
the third TsT transformation with
the shift parameter equal to ${\hg_2}$ to the torus $(\p_3, \p_1)$.

The resulting type IIB supergravity background written in string frame takes the form
\bea
\la{threepar}
ds^2_{{\rm{ str}}} &=& R^2\left[ ds^2_{{\rm{ ads}}} +
\sum_i\left(dr_i^2\+ G\, r_i^2\, d\p_i^2 \right) \+
G\, r_1^2\,r_2^2\,r_3^2\left( \sum_i \hg_i\,d\p_i\right)^2\right]\, ,\\
\nonumber
G^{-1}&=& 1 \+ \hg_3^2\,r_1^2\,r_2^2\+ \hg_1^2\,r_2^2\,r_3^2\+\hg_2^2\,r_3^2\,r_1^2\, ,
\qquad R^4\equiv 4\pi e^{\p_0}N\= \sqrt{\l}\,,\quad \a'\=1\,,\\ \nonumber
e^{2\p}&=&e^{2\p_0}\,G\, ,\\\nonumber
B^{NS}&=&R^2 G\left( \hg_3\, r_1^2\,r_2^2\,d\p_1\wedge d\p_2
\+ \hg_1\, r_2^2\,r_3^2\,d\p_2\wedge d\p_3\+\hg_2\, r_3^2\,r_1^2\,d\p_3\wedge d\p_1\right)\,,\\
\nonumber
C_2&=& -4\,R^2\,e^{-\phi_0} \, w_1\wedge \sum_i \hg_i\,d\p_i\,,\qquad \hg_i\= R^2\g_i\,,\quad
dw_1\= c_\a s_\a^3 s_\theta c_\theta\, d\a\wedge d\theta\,,\\\nonumber
C_4&=&4\,R^4\,e^{-\phi_0} \,
\left( w_4 \+ G\,w_1\wedge d\p_1\wedge d\p_2\wedge d\p_3\right)\,,\\\nonumber
F_5&=&4\,R^4\,e^{-\phi_0} \, \left( \w_{AdS_5} + G\,\w_{S^5}\right)\, ,\qquad \w_{S^5}\=
dw_1\wedge d\p_1\wedge d\p_2\wedge d\p_3\,,\quad \w_{AdS_5}\= dw_4\,,
\eea
where we used the notations from \cite{LM}: $c_\a\equiv \cos\a\,,\ s_\a\equiv\sin\a\,,\
r_1 = \cos\a\,,\ r_2=\sin\a\cos\theta$, and $\w_{S^5}$ is the volume form of unit radius $S^5$.
We also applied the rules of T-duality\footnote{Note
that the standard signs of the $B^{NS}$ and $C_2$ T-duality
rules were flipped in \cite{LM}. The T-duality rules (\ref{A7})
match the choice made in \cite{LM}.
The author thanks Oleg Lunin for a discussion of this point.}
for RR fields \cite{RARA} to find $C_2$ and
$C_4$.

The relations between the angle variables of the three-parameter deformed background
and those of \adss are still given by (\ref{currents2}), and can be used to find
the Lax representation for the model in the same way we did in section 3.

Introducing the momenta conjugated to the angle variables, the relations take the form
\bea\la{momedphi}
&&\tilde{\tilde{p}}_i \=  p_i\, .\\\nonumber
&&\ttp_1' \= \p_1'\+ \g_3\,p_2 \- \g_2\, p_3\, ,\\ \nonumber
&&\ttp_2' \= \p_2'\+ \g_1\,p_3 \- \g_3\, p_1\, ,\\ \nonumber
&&\ttp_3' \= \p_3'\+ \g_2\,p_1 \- \g_1\,p_2\, ,
\eea
where
$\g_i\= \hg_i/\sqrt{\l}$
are the deformation parameters that appear in field theory.

Twisted boundary conditions for the $U(1)$ variables $\ttp_i$ of the \adss theory
take the following form
\bea
\la{dphi33}
&&\p_i(2\pi) \- \p_i(0) \= 2\pi\, n_i\, ,\quad n_i \mbox{ are integer winding numbers}\, , \\
\nonumber
&&\ttp_1(2\pi) \- \ttp_1(0) \= 2\pi \left(n_1\+ \g_3\,J_2 \- \g_2\,J_3\right)\, ,\\ \nonumber
&&\ttp_2(2\pi) \- \ttp_2(0) \= 2\pi \left(n_2\+ \g_1\,J_3 \- \g_3\,J_1\right)\, ,\\ \nonumber
&&\ttp_3(2\pi) \- \ttp_3(0) \= 2\pi \left(n_3\+ \g_2\,J_1 \- \g_1\,J_2\right)\, .
\eea
We see that for generic values of the deformation parameters $\g_i$ all the angular variables
have nontrivial twists.

Since the background (\ref{threepar}) breaks all of the supersymmetry of \adss,
it should be dual to a nonsupersymmetric but marginal deformation of $\N=4$ SYM.\footnote{
It is unclear if the deformation is marginal for finite $N$.}
The bosonic part of the deformed YM potential should have the following form\footnote{The
deformation is not supersymmetric and cannot be written in terms of $\N=1$ superfields.}
\bea
\nonumber
V &=& \Tr\left[ |\P_1\P_2 - e^{-2i\pi\g_3}\P_2\P_1|^2 +
|\P_2\P_3 - e^{-2i\pi\g_1}\P_3\P_2|^2+
|\P_3\P_1 - e^{-2i\pi\g_2}\P_1\P_3|^2\right]\\\la{poten}
&+&\Tr\left[ \left( \left[\P_1,\,\bar{\P}_1\right] +\left[\P_2,\,\bar{\P}_2\right] +
\left[\P_3,\,\bar{\P}_3\right] \right)^2
\right]\,,
\eea
where $\P_i$ are the three holomorphic scalars of $\N=4$ SYM. The potential can be obtained
from the undeformed one by replacing the usual product $\P_i\P_j$ by the associative $*$-product
of \cite{LM}. It should be possible to obtain the fermionic part of the potential by the same
procedure, and to check if the deformation is marginal at one-loop.
It is known that the one-loop dilatation operator associated to the potential
(\ref{poten}) is described by integrable spin chains in the $su(2)_{\g_i}$\footnote{
The $su(2)_{\g_i}$ subsector coincides with the $su(2)_\g$ subsector of the \LM model.}
and $su(3)_{\g_i}$
subsectors of the deformed YM model \cite{RR}. The existence of the Lax pair representation
for the  bosonic part of the string sigma model on the
three-parameter deformed background implies that the sigma model is integrable too.
It would be interesting to find the
Bethe ansatz \cite{Fad} for the $su(3)_{\g_i}$ spin chain and string Bethe equations for
the deformed background.

The three-parameter deformed background (\ref{threepar})
has the same structure as the \LMg one. In principle
one can use $SL(2,R)_s$ transformations to generate more general solutions \cite{LM}
similar to the \LM background with complex $\b$. In fact, since a general \LM transformation
can be symbolically written as $S_\s Ts_\g T S_\s^{-1}$, where $S_\s$ denotes an $SL(2,R)_s$
transformation with a parameter $\s_s$, and $Ts_\g T$ denotes a TsT transformation with
the shift parameter $\g$, we can get a 6+2 parameter solution from \adss by performing
three consecutive STsTS transformations on each of the three tori.
Let us also mention that the step involving S-duality departs from the world sheet treatment,
see \cite{FRT} for a detailed discussion. The 6+2 parameter solution
appears to be rather complicated, in particular, it has nonvanishing
$G_{\phi_i\a}$ and $G_{\phi_i\theta}$ components of the metric. Its explicit form is given
in Appendix B. Here we present a simpler solution with
4+2 parameters corresponding to the three $\g_i$, one $\s_s$, the dilaton $\phi_0$ and the axion
$\chi_0$.  The solution
generalizes the general supersymmetry preserving  \LM deformation of \adss. It is worth noting
that the 4+2 parameter solution cannot be obtained by using just one $SL(3,R)$ transformation
of \adss. It is generated by performing the transformation
$S_\s TsT_{\g{_1}} TsT_{\g_2}TsT_{\g_3}S_{\s}^{-1}$.

To derive the solution we begin with
the $AdS_5\times S^5$ background with constant dilaton $\phi_0$ and axion $\chi$, and
perform the following $SL(2,R)$ transformation of the background
\bea
\la{sl2r}
&&\chi \+ i\, e^{-\p_0}\, \equiv\,  \tau\,\to\,\tau_{\s_s}\= {\tau\ov 1 \-\s_s\, \tau}\,,\qquad
B^{NS}\,\to\, B^{NS}_{\s_s}\= B^{NS} \- \s_s\, C_{2}\,,\qquad\\\nonumber
&&C_4\,\to\, C_4^{\s_s}\= C_4 \- {1\ov 2}\,\s_s\, C_{2}\wedge C_2\, .
\eea
The Einstein frame metric and the two-form $C_2$ remain invariant under this transformation.
It is worth mentioning that the parameter $\s_s$ is not equal to
the parameter $\s$ used in \cite{LM}.

Then we perform the same chain of TsT transformations we used to generate the
three-parameter deformed background (\ref{threepar}). Finally, we perform
the inverse $SL(2,R)$ transformation with $\s_s$ replace by $-\s_s$ in (\ref{sl2r}).

The resulting 4+2 parameter type IIB supergravity background
written in string frame takes the form
\bea
\la{fourpar}
ds^2_{{\rm{ str}}} &=& R^2\,H^{1/2}\left[ ds^2_{{\rm{ ads}}} +
\sum_i\left(dr_i^2\+ G\, r_i^2\, d\p_i^2 \right) \+
G\, r_1^2\,r_2^2\,r_3^2\left| \sum_i \tb_i\,d\p_i\right|^2\right]\, ,
\\ \nonumber
e^{2\p}&=&e^{2\p_0}\,G\,H^2\, ,\qquad \chi\=\chi_0 \+ e^{-\phi_0}\,H^{-1}Q\,,
\\\nonumber
B^{NS}&=&R^2\left( G\, w_B - 4\,w_1\wedge \sum_i \ts_i\,d\p_i
\right)
\,,\\
\nonumber
C_2&=& R^2\,\left( G\, w_C
 -4\,w_1\wedge \sum_i (e^{-\p_0}\tg_i\+ \chi_0\ts_i )\,d\p_i \right)\,,
\\\nonumber
C_4&=&4\,R^4\,e^{-\phi_0} \,
\left( w_4 \+ G(H-e^{\p_0}\chi_0 Q)\,w_1\wedge d\p_1\wedge d\p_2\wedge d\p_3\right)\,,\\\nonumber
F_5&=&4\,R^4\,e^{-\phi_0} \, \left( \w_{AdS_5} + G\,\w_{S^5}\right)\, .
\eea
Here the functions $G$, $H$, and $Q$, and the parameters
$\tb_i$, $\tg_i$ and $\ts_i$ are defined as follows
\bea
\la{GHQ}
G^{-1}&=& 1 \+ |\tb_3|^2\,r_1^2\,r_2^2\+ |\tb_1|^2\,r_2^2\,r_3^2\+|\tb_2|^2\,r_3^2\,r_1^2\, ,
\\
\nonumber
H&=& 1 \+ \ts_3^2\,r_1^2\,r_2^2\+ \ts_1^2\,r_2^2\,r_3^2\+\ts_2^2\,r_3^2\,r_1^2\, ,
\\
\nonumber
Q&=&\ts_3\tg_3\,r_1^2\,r_2^2\+ \ts_1\tg_1\,r_2^2\,r_3^2\+\ts_2\tg_2\,r_3^2\,r_1^2\, ,
\\\nonumber
w_B&=&\tg_3\, r_1^2\,r_2^2\,d\p_1\wedge d\p_2
\+ \tg_1\, r_2^2\,r_3^2\,d\p_2\wedge d\p_3\+\tg_2\, r_3^2\,r_1^2\,d\p_3\wedge d\p_1
\,,
\\\nonumber
w_C&=&\chi_0\,w_B \- e^{-\p_0}\left(\ts_3\, r_1^2\,r_2^2\,d\p_1\wedge d\p_2
\+ \ts_1\, r_2^2\,r_3^2\,d\p_2\wedge d\p_3\+
\ts_2\, r_3^2\,r_1^2\,d\p_3\wedge d\p_1\right)
\,,\\\nonumber
\tg_i &=& R^2\g_i (1 - \chi_0\s_s) = \hg_i (1 - \chi_0\s_s)\,,\quad
\ts_i= R^2 e^{-\phi_0}\g_i\s_s = \hg_i e^{-\phi_0}\s_s\,,\quad
\tb_i = \tg_i\- i \ts_i\,.
\eea
It is easy to see that if all $\g_i$ are equal to each other then the background
coincides with the \LM one provided the parameter $\s$ in \cite{LM} is related to
$\s_s$ as follows
\bea
\nonumber
\s\=\s_s \g\,,\quad \g_i \=\g\,.
\eea
The 4+2 parameter background should be dual to a nonsupersymmetric marginal
(at the large $N$ limit)
deformation of $\N=4$ SYM. It would be interesting to find
this dual nonsupersymmetric YM model.

\section{Conclusion}

In this paper we have discussed the TsT transformation of the \adss background, and
shown how it can be used to generate the \LM supergravity solution in the case of
the real deformation parameter $\g$. We have used the TsT transformation to find
the relation between the angle variables of \adss, and the angle variables of the \LMg
background, and used the relation to derive a local and periodic Lax representation
for the \LMg model. The existence of the Lax pair implies the integrability
of the (bosonic part of) string sigma model on the \LMg background.

It is clear that it should be possible to apply the TsT transformation
to the Green-Schwarz $\k$-symmetric
superstring action on \adss \cite{MT} to generate the \LMg
background with all the supergravity fields included. To this end one can try to use
the rules of T-duality formulated in \cite{RK} for the Green-Schwarz superstring.
Then, the approach used in section 3 of this paper should lead to a local and periodic
Lax representation for the complete Green-Schwarz sigma model on the \LMg supergravity
background.

It would be interesting to use the Lax representation to
analyze the properties of the monodromy matrix and derive
the string Bethe equations for the \LM model analogous to those derived for strings
on $AdS_5\times S^5$ in \cite{KMMZ,KMMZ2,AF}. The string Bethe equations could be then
compared with the thermodynamic limit of the Bethe equations for the \LMg $\N=4$
SYM theory \cite{RR,BECH}. It has been already done for the simplest $su(2)_\g$ case in
\cite{FRT}.

As another interesting application of the TsT transformation we generated
the three-parameter regular supergravity background
by using a chain of TsT transformations applied to different
tori of \adss. This background is expected to be dual to a
nonsupersymmetric marginal deformation
of $\N=4$ SYM theory. It should be possible to perform a detailed analysis of the
three-parameter background, and the dual conformal field theory.

We also derived a $6+2$ parameter deformation of \adss by applying a chain of
STsTS transformations to the three tori of $S^5$. The type IIB supergravity solution
is nonsupersymmetric, and it is important to check if it is perturbatively stable.
Since it depends on the continuous deformation parameters one might expect that the
background is stable at least for small values of the parameters.
It also would be interesting to
find  the nonsupersymmetric conformal
deformation of $\N=4$ SYM model dual to this $6+2$ parameter background.

It is of interest to generate other multi-parameter regular backgrounds by using
a chain of the TsT and STsTS transformations applied to \adss, and other supergravity
backgrounds with $U(1)^3$ isometry. In particular, one can consider nonsupersymmetric marginal
deformations of theories based on toric manifolds \cite{Kleb,Toric,Benv}. One can also use the
TsT transformations to derive nonsupersymmetric deformations of supergravity
backgrounds dual to nonconformal field theories such as the Klebanov-Strassler background
\cite{KlSt}.

In general, the \LM type backgrounds and the TsT (STsTS) transformation
have many interesting
properties which are worth studying.

\setcounter{section}{0}
\setcounter{equation}{0}
\setcounter{footnote}{0}
\vspace{0.2in}
{\bf Acknowledgments}
\vspace{0.1in}

The author is grateful to Gleb Arutyunov, Oleg Lunin, Radu Roiban and Arkady Tseytlin
for useful correspondence and discussions.


\appendix{T-duality Transformation}
\renewcommand{\theequation}{A.\arabic{equation}}
\setcounter{equation}{0}

In this Appendix we present the T-duality transformation \cite{Buscher} in the form
used in the paper. We start with the following string theory action
\bea
\la{Aa1}
S = -{\sqrt{\lambda}\over 2}\int\, d\tau {d\s\over
2\pi} \left[ \g^{\a\b}\partial_\a X^M\partial_\b X^N\, G_{MN}(X^i) -
\e^{\a\b}\partial_\a X^M\partial_\b X^N\, B_{MN}(X^i)\right]\, .
\eea
Here $\e^{01}=1$, $M=1,2,3\dots$, $i=2,3\dots$, and the background fields $G_{MN}$ and
$B_{MN}$ do not depend on $X^1$. The action can be represented in the following equivalent form
\bea
\la{Aa2}
&&S = -\sqrt{\lambda}\int\, d\tau {d\s\over
2\pi} \left[ p^\a\left(\pa_\a X^M{G_{1M}\ov G_{11}}
- \g_{\a\b}\e^{\b\r}\pa_\r X^M{B_{1M}\ov G_{11}}\right)\right.
-{1\ov 2\, G_{11}}\g_{\a\b}\,p^\a p^\b \\ \nonumber
&&\left. +{1\ov 2} \g^{\a\b}\partial_\a X^i\partial_\b X^j
\left( G_{ij} -{G_{1i}G_{1j}-B_{1i}B_{1j}\ov G_{11}}\right)-
{1\ov 2} \e^{\a\b}\partial_\a X^i\partial_\b X^j
\left( B_{ij} -{G_{1i}B_{1j}-B_{1i}G_{1j}\ov G_{11}}\right)\right]\, .
\eea
Indeed, variating with respect to $p^\a$, one gets the following equation of motion for $p^\a$
\bea
\la{A3}
p^\a \= \g^{\a\b}\pa_\b X^M G_{1M} \- \e^{\a\b}\pa_\b X^M B_{1M}\, .
\eea
Substituting (\ref{A3}) into (\ref{Aa2}) and using the identity
$\e^{\a\g}\g_{\g\r}\e^{\r\b} \= \g^{\a\b}$, we reproduce the action (\ref{Aa1}).

On the other hand, variating (\ref{Aa2}) with respect to $X^1$ gives
\bea
\la{A4}
\pa_\a\, p^\a\= 0\, .
\eea
The general solution to this equation can be written in the form
\bea\la{A5}
p^\a\= \e^{\a\b}\pa_\b \tilde{X}^1\, ,
\eea
where $\tilde{X}^1$ is the scalar T-dual to $X^1$. Substituting (\ref{A5}) into the action
(\ref{Aa2}), we obtain the following T-dual action
\bea
\la{A6}
\tilde{S} = -{\sqrt{\lambda}\over 2}\int\, d\tau {d\s\over
2\pi} \left[ \g^{\a\b}\partial_\a \tilde{X}^M\partial_\b \tilde{X}^N\, \tilde{G}_{MN} -
\e^{\a\b}\partial_\a \tilde{X}^M\partial_\b \tilde{X}^N\, \tilde{B}_{MN}\right]\, ,
\eea
where
\bea\la{A7}
&&\tilde{G}_{11} \= {1\ov G_{11}}\, , \quad
\tilde{G}_{ij} \= G_{ij} -{G_{1i}G_{1j}-B_{1i}B_{1j}\ov G_{11}}\, ,
\quad \tilde{G}_{1i}\= {B_{1i}\ov G_{11}}\, ,
\\ \nonumber
&&
\tilde{B}_{ij} \= B_{ij} -{G_{1i}B_{1j}-B_{1i}G_{1j}\ov G_{11}}\, ,
\quad \tilde{B}_{1i}\= {G_{1i}\ov G_{11}}\, ,
\\ \nonumber
&&\e^{\a\b}\pa_\b \tilde{X}^1\= \g^{\a\b}\pa_\b X^M G_{1M} \- \e^{\a\b}\pa_\b X^M B_{1M}\, ,
\quad \tilde{X}^i \= X^i\, .
\eea


\appendix{6+2 parameter background}
\renewcommand{\theequation}{B.\arabic{equation}}
\setcounter{equation}{0}

The 6+2 parameter solution can be obtained by performing the transformation
$$S_{\s_1^s} TsT_{\g{_1}}S_{\s_2^s-\s_1^s} TsT_{\g_2}S_{\s_3^s-\s_2^s}TsT_{\g_3}S_{-\s_3^s}\,,$$
that is a chain of the three consecutive STsTS transformations on each of the three tori.

The resulting 6+2 parameter type IIB supergravity background
written in string frame takes the form
\bea
\la{sixpar}
ds^2_{{\rm{ str}}} &=& R^2\,H^{1/2}\Big[ ds^2_{{\rm{ ads}}} +
\sum_i\left(dr_i^2\+ G\, r_i^2\, d\p_i^2 \right) \+
G\, r_1^2\,r_2^2\,r_3^2\Big| \sum_i \tb_i\,d\p_i\Big|^2
\\ \nonumber
&+& 8\,G\,w_1\left( (\hg_3\ts_2 -\hg_2\ts_3) r_1^2 d\p_1\+
(\hg_1\ts_3 -\hg_3\ts_1) r_2^2 d\p_2  \+
(\hg_2\ts_1 -\hg_1\ts_2) r_3^2 d\p_3 \right)
\\ \nonumber
&+& 16\,G\,w_1^2\left( (\hg_3\ts_2 -\hg_2\ts_3)^2 r_1^2  \+
(\hg_1\ts_3 -\hg_3\ts_1)^2 r_2^2   \+
(\hg_2\ts_1 -\hg_1\ts_2)^2 r_3^2 \right)
\Big]\, ,
\\ \nonumber
e^{2\p}&=&e^{2\p_0}\,G\,H^2\, ,\qquad \chi\=\chi_0 \+ e^{-\phi_0}\,H^{-1}Q\,,
\\\nonumber
B^{NS}&=&R^2\,G\, \left( w_B - 4\,w_1\wedge A_B
\right)
\,,\\
\nonumber
C_2&=& R^2\,G\, \left( w_C
 -4\,e^{-\p_0}\,w_1\wedge A_C \right)\,,
\\\nonumber
C_4&=&4\,R^4\,e^{-\phi_0} \,
\left( w_4 \+ G(H-e^{\p_0}\chi_0 Q)\,w_1\wedge d\p_1\wedge d\p_2\wedge d\p_3\right)\,,\\\nonumber
F_5&=&4\,R^4\,e^{-\phi_0} \, \left( \w_{AdS_5} + G\,\w_{S^5}\right)\, .
\eea
Here the functions $G$, $H$, and $Q$, and the parameters
$\tb_i$, $\tg_i$ and $\ts_i$ are defined as follows
\bea
\la{GHQg}
G^{-1}&=& 1 \+ |\tb_3|^2\,r_1^2\,r_2^2\+ |\tb_1|^2\,r_2^2\,r_3^2\+|\tb_2|^2\,r_3^2\,r_1^2
\\ \nonumber
&+& r_1^2r_2^2r_3^2\left( (\hg_3\ts_2 -\hg_2\ts_3)^2 r_1^2  \+
(\hg_1\ts_3 -\hg_3\ts_1)^2 r_2^2   \+
(\hg_2\ts_1 -\hg_1\ts_2)^2 r_3^2 \right)\, ,
\\
\nonumber
H&=& 1 \+ \ts_3^2\,r_1^2\,r_2^2\+ \ts_1^2\,r_2^2\,r_3^2\+\ts_2^2\,r_3^2\,r_1^2\, ,
\\
\nonumber
Q&=&\ts_3\tg_3\,r_1^2\,r_2^2\+ \ts_1\tg_1\,r_2^2\,r_3^2\+\ts_2\tg_2\,r_3^2\,r_1^2\, ,
\\\nonumber
w_B&=&\left[ \tg_3\, r_1^2\,r_2^2 \+ r_1^2r_2^2r_3^2\left(\ts_1(\hg_3\ts_1 -\hg_1\ts_3) r_2^2\+
\ts_2(\hg_3\ts_2 -\hg_2\ts_3) r_1^2\right)
\right]d\p_1\wedge d\p_2\\\nonumber
&\+& \left[ \tg_1\, r_2^2\,r_3^2 \+r_1^2r_2^2r_3^2\left(\ts_2(\hg_1\ts_2 -\hg_2\ts_1) r_3^2\+
\ts_3(\hg_1\ts_3 -\hg_3\ts_1) r_2^2\right)
\right]d\p_2\wedge d\p_3
\\\nonumber
&\+&\left[ \tg_2\, r_3^2\,r_1^2 \+r_1^2r_2^2r_3^2\left(\ts_3(\hg_2\ts_3 -\hg_3\ts_2) r_1^2\+
\ts_1(\hg_2\ts_1 -\hg_1\ts_2) r_3^2\right)
\right]d\p_3\wedge d\p_1
\,,
\\\nonumber
A_B&=&\left[ \ts_1 \+ \ts_1 (\tg_1^2+\ts_1^2)r_2^2r_3^2 \+\ts_2(\tg_1\tg_2
+\ts_1\ts_2)r_3^2r_1^2 \+
\ts_3(\tg_1\tg_3+\ts_1\ts_3)r_1^2r_2^2\right]\, d\p_1\\\nonumber
&+&\left[ \ts_2 \+ \ts_2 (\tg_2^2+\ts_2^2)r_3^2r_1^2 \+\ts_1(\tg_1\tg_2+\ts_1\ts_2)r_2^2r_3^2 \+
\ts_3(\tg_2\tg_3+\ts_2\ts_3)r_1^2r_2^2\right]\, d\p_2\\\nonumber
&+&\left[ \ts_3 \+ \ts_3 (\tg_3^2+\ts_3^2)r_1^2r_2^2 \+\ts_1(\tg_1\tg_3+\ts_1\ts_3)r_2^2r_3^2 \+
\ts_2(\tg_2\tg_3+\ts_2\ts_3)r_3^2r_1^2\right]\, d\p_3
\,,
\\\nonumber
w_C&=&\left[ (\chi\tg_3-e^{-\p_0}\ts_3)r_1^2r_2^2 + e^{-\p_0}r_1^2r_2^2r_3^2
\left(\hg_1(\hg_3\ts_1 -\hg_1\ts_3) r_2^2+
\hg_2(\hg_3\ts_2 -\hg_2\ts_3) r_1^2\right)
\right]d\p_1\wedge d\p_2\\\nonumber
&+& \left[ (\chi\tg_1-e^{-\p_0}\ts_1)r_2^2r_3^2 +e^{-\p_0}r_1^2r_2^2r_3^2
\left(\hg_2(\hg_1\ts_2 -\hg_2\ts_1) r_3^2+
\hg_3(\hg_1\ts_3 -\hg_3\ts_1) r_2^2\right)
\right]d\p_2\wedge d\p_3
\\\nonumber
&\+&\left[(\chi\tg_2-e^{-\p_0}\ts_2)r_3^2r_1^2 +e^{-\p_0}r_1^2r_2^2r_3^2
\left(\hg_3(\hg_2\ts_3 -\hg_3\ts_2) r_1^2+
\hg_1(\hg_2\ts_1 -\hg_1\ts_2) r_3^2\right)
\right]d\p_3\wedge d\p_1
\,,
\\\nonumber
A_C&=&\left[ \hg_1 \+ \hg_1 (\tg_1^2+\ts_1^2)r_2^2r_3^2 \+\hg_2(\tg_1\tg_2
+\ts_1\ts_2)r_3^2r_1^2 \+
\hg_3(\tg_1\tg_3+\ts_1\ts_3)r_1^2r_2^2\right]\, d\p_1\\\nonumber
&+&\left[ \hg_2 \+ \hg_2 (\tg_2^2+\ts_2^2)r_3^2r_1^2 \+\hg_1(\tg_1\tg_2+\ts_1\ts_2)r_2^2r_3^2 \+
\hg_3(\tg_2\tg_3+\ts_2\ts_3)r_1^2r_2^2\right]\, d\p_2\\\nonumber
&+&\left[ \hg_3 \+ \hg_3 (\tg_3^2+\ts_3^2)r_1^2r_2^2 \+\hg_1(\tg_1\tg_3+\ts_1\ts_3)r_2^2r_3^2 \+
\hg_2(\tg_2\tg_3+\ts_2\ts_3)r_3^2r_1^2\right]\, d\p_3
\,,\\\nonumber
\tg_i &=& R^2\g_i (1 - \chi_0\s^s_i) = \hg_i (1 - \chi_0\s^s_i)\,,\quad
\ts_i= R^2 e^{-\phi_0}\g_i\s^s_i = \hg_i e^{-\phi_0}\s^s_i\,,\quad
\tb_i = \tg_i\- i \ts_i\,.
\eea
The self-dual five-form $F_5$ is expressed in terms of $C_4$, $C_2$ and $B$ as follows
$$F_5 \= dC_4\- dB^{NS}\wedge C_2\,.$$

\end{document}